# Digital Labor Platforms as Infrastructure of Extractivism

**Julian Posada**
**Yale University**
**julian.posada@yale.edu**
**ORCID 0000-0002-3285-6503**

## ABSTRACT

This chapter examines how digital platforms, as infrastructures of extractivism in contemporary artificial intelligence, enable corporations to appropriate data and labour while concentrating value in core regions such as Silicon Valley. It proposes platform extractivism as a framework linking commodification, epistemic dominance, and reproduction through territory. Drawing on digital labour studies, Latin American critical thought, and mixed-methods research with Venezuelan and Colombian data workers, the chapter discusses how AI depends on precarious platform labour to produce, clean, annotate, and maintain data. It further shows how annotation systems normalize culturally specific categories, especially US norms, as universal knowledge, reproducing colonial hierarchies. Finally, it highlights how data work is sustained by unpaid reproductive labour, community networks, and material conditions that make workers available for platform labour. Platform extractivism thus reveals AI as a labour-intensive and territorially embedded political-economic system under capitalism that perpetuates legacies of coloniality.

## 1. INTRODUCTION

Many workers experience so-called artificial intelligence (AI) as a technology that *takes*—from artists who encounter their work reflected in the outputs of image-generating models to writers who find their texts used to train large language models without consent or compensation. Machine learning, which Brian Cantwell Smith (2019) described as the 'second wave' of AI, distinguished from the symbolic AI that dominated the field during the second half of the 20th century, has been successful not solely because of the commonly attributed advances in computational power and the availability of large datasets but also because of extensive human labour. Machine learning depends on vast quantities of data to train models and produce outputs deemed increasingly accurate, yet this dependence is inseparable from the social relations through which such data are produced, collected, and maintained.

In this chapter, I examine the digital platform as a form of infrastructure for extractivism and as a central driver of the provision of both data and labour that have enabled the development of second-wave, machine learning-based AI (hereafter called just 'AI' for simplicity). Although digital platforms do facilitate the extraction of data, from artistic production to everyday interactions with computational systems, the term 'extraction' alone does not fully capture the dynamics at play. The concept of extractivism looks beyond the action of extraction to study the 'transnational system of accumulation, labor exploitation, and power concentration' that surrounds it (Posada, 2026). Understood as a system, extractivism encompasses the social, economic, and political conditions that enable and sustain acts of extraction. From this perspective, the computing industry's demand for data situates digital platforms within legal, political, and economic frameworks that echo those historically associated with other extractive industries.

Conceptualizing platforms through the lens of extractivism enables three analytical moves. First, it situates platforms within the longer historical trajectory of extractivism characterized by core–periphery relations. Second, it links platforms to questions of colonialism and coloniality that have become increasingly prominent in critical scholarship. Third, it foregrounds the inseparable relationships among data extraction, data production, and labour. These dimensions are deeply intertwined: historical extractivism has always relied on labour exploitation and power concentration, both of which remain central to the study of contemporary digital platforms.

Within this context, I propose a framework for analysing platform extractivism organized around three dimensions. The first concerns commodification: the transformation of both data and labour into economic assets within AI development, alongside data extraction itself. The second dimension is epistemological, addressing how AI development entails the imposition of particular forms of knowledge and worldviews through processes of data extraction. The third-dimension concerns reproduction and its relationship to territory, highlighting how extractive dynamics are sustained and spatially organized over time.

This chapter draws primarily on the literature in digital labour studies and Latin American critical thought, as well as on my own empirical research conducted between 2018 and 2025 on the working conditions and lived experiences of data workers in Venezuela and Colombia, with particular emphasis on the former. This research employed a mixed-methods approach, combining qualitative and quantitative techniques. Specifically, I conducted in-depth interviews with platform workers and managers, surveyed more than 600 workers, carried out web traffic analyses of platform interfaces for both clients and workers, and performed discourse analysis of the annotation guidelines and instructions provided to workers.

## 2. EXTRACTIVISM AND LABOUR COMMODIFICATION

Extractivism does not refer solely to the act of extracting specific resources or commodities but rather to the system that enables, organizes, and sustains such extraction. The term gained particular prominence in the mid-twentieth century, when institutions such as the World Bank (2025) promoted the development of so-called extractive industries in Latin America as a pathway to economic growth, integrating the region into global supply chains oriented toward the export of raw materials. Extraction has long been a defining feature of this region's economic history. In the early twenty-first century, a renewed wave of scholarly debate emerged as several Latin American countries intensified resource extraction to finance social programmes, prompting analysts to introduce the concept of neo-extractivism to distinguish these policies from earlier forms of extraction (Gudynas, 2020; Centro Andino de Acción Popular y Centro Latinoamericano de Ecología Social, 2009).

This historical context is central to the empirical component of my research and underpins many of the arguments developed in this chapter. Critics of neo-extractivism, including Maristella Svampa (2012; 2015), have examined cases such as Venezuela under the government of Hugo Chávez. In this instance, heavy reliance on oil exports contributed to economic vulnerability, culminating in a severe crisis in the mid-2010s following declining oil prices, among other factors, including international sanctions. The resulting crisis, marked at one point by one of the highest inflation rates globally, led many Venezuelans to seek online employment that paid in US dollars, a currency considerably more stable than the local *bolívar* (Roy & Cheatham, 2024; Corrales, 2022; Instituto de Investigaciones Económicas y Sociales, 2024). Though Venezuela represents a specific case, it illustrates some of the broader consequences of extractivist policies, including environmental degradation and profound transformations in the living and working conditions of populations residing in or dependent upon extractive economies.

At its core, extractivism describes a system of resource appropriation structured through relations between periphery and core, relying on centralized power—often exercised by monopolistic industries—to facilitate capital accumulation (Chagnon et al. 2022; Ye et al., 2020). The use of the terms 'core' and 'periphery' stems from the long tradition of dependency theory in Latin America, which critically examined patterns of economic development in the region and the persistent power asymmetries and unequal wealth distributions shaping its relationships with wealthier states and institutions elsewhere (Mahoney & Rodríguez-Franco, 2016). As a systemic critique, extractivism foregrounds processes of depletion at sites of extraction. Although the concept has historically been associated with the environmental harms produced by natural resource extraction, depletion extends beyond ecosystems to impact human life, particularly the communities and workers whose labour sustains extractive systems. Importantly, extractivism does not supersede capitalism or colonialism; rather, it names a specific configuration of capitalist accumulation shaped by colonial legacies, including power asymmetries, logics of accumulation, structural inequality, and enduring core–periphery relations in global commodity chains (Posada, 2026).

The notion of the commodity is particularly salient for understanding contemporary AI development. Data—the central input of second-wave AI over the past several decades—should not be understood as a commodity in a strict sense. Nevertheless, it is routinely acquired, exchanged, and monetized by corporations. This contradiction underpins my use of the term data extractivism, which is closely related to the concept of platform extractivism mobilized in this chapter. Even as data constitutes a fictitious commodity following Karl Polanyi (2001 [1944]), it is nonetheless commodified within markets organized by data platforms that generate, annotate, and validate algorithmic outputs or, in other words, data work (Miceli & Posada, 2022; Tubaro et al., 2020). These transactions form a central mechanism of the extractivist dynamics under examination here.

Like other forms of extractivism, platform-mediated extractivism depends fundamentally on labour. In the context of AI, data extraction is inseparable from data transformation—that is, from processes of data production. The concept of platform extractivism builds on scholarship in digital labour studies that has documented how remote online labour, often performed in the majority world, supports the production of data and services for firms based in wealthier economies (Braesemann et al., 2022; Tubaro & Casilli, 2022). Platforms, understood as digital infrastructures that 'facilitate and shape personalized interactions between end users and complementors, organized through the systematic collection, algorithmic processing, monetization, and circulation of data' (Poell et al., 2019), occupy a central position within contemporary extractivist systems. Much like earlier infrastructures, ranging from mines to shipping ports, that enabled the extraction and movement of resources from periphery to core, platforms function as key logistical and organizational mechanisms for data extraction. Although platforms ostensibly connect companies to a global user base in order to provide services, such as communication and e-commerce, their significance for AI development lies in their capacity to collect large-scale data and grant access to an expanded, deregulated, and readily exploitable labour force. This labour is integral to the production, annotation, and refinement of data upon which machine learning systems depend.

Framing platforms as infrastructures of data extractivism extends these discussions by emphasizing labour conditions as well as epistemological and territorial dimensions. These dimensions are essential for understanding AI development as a system of power concentration structured through core-periphery relations, one that exceeds the boundaries of individual platforms and has lasting epistemological consequences for the data upon which AI systems are built.

## 3. EPISTEMIC DOMINANCE

Understanding extractivism as a system of structural power imbalance also requires attention to its epistemological dimensions. As a system rooted in coloniality, extractivism has historically entailed not only the removal of wealth and resources but also the imposition of legal, social, and epistemic orders designed to guarantee and legitimize wealth accumulation (Ricaurte Quijano, 2019; 2023). Extractivism is therefore not limited to the extraction of material resources but also encompasses the extraction, circulation, and imposition of ideas; this is particularly relevant in the case of data, which does not fit neatly into traditional material categories. As Silvia Rivera Cusicanqui (2012) observes, 'ideas run, like rivers, from the South to the North and are transformed into tributaries in major waves of thought', a simile she uses to describe how scholars in European and North American institutions appropriate ideas originating in Latin America without attribution. This comparison is particularly apt for understanding how digital platforms enable the flow of ideas encoded as data, channelling them toward centres of power, where they are transformed into the foundational components of contemporary AI systems. Because extraction and imposition are inseparable, data production processes also involve the normalization of particular forms of knowledge and the erasure of others.

In collaborative work with Milagros Miceli, I examined the annotation guidelines provided to data workers across three digital platforms and one business process outsourcing firm operating in Latin America (Miceli & Posada, 2022). Using dispositive analysis (a form of discourse analysis drawing on work by Michel Foucault), we identified how these instructions shaped workers' cognitive and interpretive frameworks. Specifically, workers were encouraged to reproduce cultural categories and norms largely derived from the United States, embedding them into datasets as 'ground truth'. Such norms included, for example, naming conventions for objects and systems of social classification, such as racial categories, that are culturally specific to the US context but treated as universal. Workers' precarious living and working conditions further constrained their ability to contest these epistemic impositions, as deviation from prescribed guidelines exposed them to algorithmic surveillance, arbitrary disciplinary measures, and the risk of account deactivation—forms of control characteristic of the gig economy (Gandini, 2019; Woodcock & Graham, 2020).

These practices exemplify what I describe as epistemic dominance within platform extractivism (Posada, 2026). They demonstrate how the power asymmetries central to extractivism extend beyond material relations to encompass cognition, knowledge production, and meaning making. In this sense, platform extractivism reproduces colonial legacies that remain deeply embedded in Latin American contexts. An ideological infrastructure legitimizes these processes while rendering workers and their contributions largely invisible. To address this dimension, I turn to the apparent contradiction between the techno-libertarian ethos long associated with the technology sector (often described as the 'Californian ideology', with its emphasis on deregulation and freedom) and the industry's exploitative labour practices abroad, alongside the increasingly explicit alignment of major technology actors with far-right political movements.

This contradiction can be understood through the enduring liberal traditions within capitalism and coloniality that, as Lisa Lowe (2015) argues, have historically justified economic exploitation and domination in sites of extraction while simultaneously extending ideals of freedom and emancipation to those positioned at the imperial centre. Within contemporary platform extractivism, the promise of techno-libertarian freedom is revealed to be profoundly uneven: it sustains hegemonic regimes of commodity extraction and epistemic imposition rather than offering universal emancipation.

Given these historical continuities and the epistemic dominance embedded in data production and extraction, it is essential to foreground ideological and epistemological dimensions in analyses of digital labour. The domain of ideas remains central: not only in processes of skill formation and deskilling, but also in the ways information systems and computational infrastructures are designed, trained, and operationalized. This perspective connects labour and production to broader analytical frameworks concerned with the circulation of knowledge across territories. Ultimately, because labour is socially and spatially embedded, its effects extend beyond the immediate sites of production into the territories and epistemic orders shaped by platform-mediated extractivism.

## 4. SOCIAL REPRODUCTION AND TERRITORY

Digital labour studies have extensively examined the relationship between platform labour and social reproductive labour (Bhutani Vij, 2023; Jarrett, 2014; Mezzadri, 2023; Wallis, 2021). This relationship is particularly salient in the context of data work, much of which, when outsourced through platforms, is performed from workers' homes. In my research, the centrality of social reproduction became evident through the observation that most data workers depended on the reproductive labour of family members to participate in platform work (Posada, 2022). Among the Venezuelan data workers in my study, not a single interviewee or survey respondent lived alone: all resided with family members. In many cases, these workers were the primary economic providers for their households, whereas other family members contributed unpaid labour in the form of domestic work and care, including cooking, childcare, and elder care. This redistribution of reproductive labour was essential to sustaining the workers' availability for paid data work.

The concept of social reproduction—originating from Karl Marx (1887 [1867]) and later developed by feminist scholars, most notably those associated with the Wages for Housework movement and beyond (Dalla Costa & James, 1975; Federici, 2021)—highlights how reproductive labour increasingly functions as a value-producing practice. This scholarship has shown how capitalism continually blurs the boundaries between productive and reproductive time (Mezzadri, 2023), a dynamic echoed in workerist notions of the social factory (Tronti, 2013 [1962]). Within the longer historical continuum of piecework, these blurred boundaries are not new; however, digital platforms have made it easier to displace them into the domestic sphere (Dubal, 2020; Gray & Suri, 2019). One of the central findings of this research, therefore, is that the outsourcing of data production through platforms is enabled not only by labour conditions and organizational arrangements but also by the availability of social reproductive labour that sustains this workforce outside formal employment institutions (Posada, 2022; Wood et al., 2019). In other words, although low wages and precarious socioeconomic conditions in sites of data extraction are critical factors in digital labour, it is unpaid reproductive labour that renders workers continuously available in front of their computers, thereby making this labour especially inexpensive.

Expanding the analytical lens beyond workers as isolated individuals toward their embeddedness within broader social and material environments reveals increasingly complex interdependencies. For instance, when examining workers' reliance on neighbours, I found that in Venezuela, amid prolonged economic and political crises, the retrenchment of state and institutional support, and the compounded effects of the COVID-19 pandemic, workers depended not only on close social ties, such as family and friends, but also on local community networks. Neighbours played a crucial role in helping workers access and manage essential resources for subsistence, including food and water. This observation drew attention to the role of non-human actors—such as water, electricity, and clean air—which are often taken for granted but become acutely visible when scarce. Extending the earlier logic, if there is no data extraction without workers, and if there are no workers without social reproductive labour, then there are likewise no workers without the environmental conditions and resources that sustain life itself.

To conceptualize this broader set of relationships, I turn to the Latin American notion of territory, a multifaceted and contested concept that emerged from political struggles and activist movements in the region and has been further elaborated by critical scholars. Silvia Rivera Cusicanqui (2018) defines territory as ‘a very complex and infinitely varied conjunction of living and palpitating beings’, whereas Arturo Escobar (2018) describes it as ‘a shorthand for the system of relations whose continuous re-enactment re-creates a community’. These relational configurations—often extended in post-anthropocentric approaches to include elements such as land and water (la_jes, 2019)—are what ultimately make production processes possible. Thus, territory offers a framework for understanding the interdependence between workers and their social and material surroundings.

When production is organized in extractivist terms, privileging value extraction over the reproduction of life, territories are degraded. In classical forms of extractivism, such as mining or fossil fuel extraction, this degradation manifests as environmental destruction. In data extractivism, its effects are inscribed on bodies, both physically, through documented health harms experienced by platform workers, and psychologically, as seen in the mental health impacts reported by content moderators and other data workers. It is therefore not coincidental that scholars and activists such as Julieta Paredes Carvajal (2012), Lorena Cabnal (2010), and Paola Ricaurte Quijano (2023) emphasize the body as the first territory. Under extractivist regimes, this primary territory is often the first to bear the costs of extraction.

## 5. CONCLUSION

The contemporary development of AI, particularly as led by large technology corporations, follows long-standing logics of extractivism deeply embedded in the historical entanglement of capitalism and colonialism—two systems as distinct yet inseparable as the strands of a double helix. From this perspective, AI can be understood as an extractivist technology. It functions through processes of appropriation and concentration, extracting value while largely without meaningfully redistributing its wealth or benefits to society at large. In particular, those who have provided the data that sustains contemporary AI systems (artists, writers, users, and especially data workers whose labour has made these systems possible over decades) are rarely compensated in proportion to their contributions.

Contemporary machine learning is often framed as the natural outcome of increased computational power and the availability of vast datasets. However, this narrative obscures a more fundamental layer of analysis. When examined through a socioeconomic lens, it becomes evident that labour is central to these developments. The production, cleaning, labelling, and maintenance of data are not incidental processes but foundational forms of work without which machine learning systems could not function.

Digital platforms have played a crucial role in this process. They have served as the primary infrastructural framework through which corporations capture and appropriate data while simultaneously organizing networks of labour exchange. Through platform-mediated data work, raw information is moulded into forms legible and valuable to the computing industry. This mode of platform extractivism closely resembles earlier forms of extractivism: it is oriented toward wealth extraction, characterized by asymmetrical power relations, and marked by the concentration of value in core regions. In this case, that core is often symbolically and materially represented by Silicon Valley in the United States. The wealth extracted consists of the datafied commodification of knowledge and skill drawn not only from data workers but also from a broad range of other contributors, including users, artists, and writers.

Employing extractivism as an analytical lens within the historical intertwining of capitalism and colonialism enables an examination of data work across both the epistemological and territorial dimensions. Epistemologically, extraction and imposition operate simultaneously. Decisions about which forms of knowledge are selected, generated, and reproduced are inseparable from the ideologies that justify and shape the design, development, and deployment of platforms. Certain knowledges are amplified, while others are marginalized or rendered invisible, reinforcing existing hierarchies of power.

The territorial dimension extends the analysis beyond the individual data or platform worker to include the territories from which labour emerges. Drawing on Rivera Cusicanqui's notion of 'palpitating beings', workers are understood not as isolated units of labour but as embedded within complex social and material ecologies. There is no data without workers, no workers without territory, and no territory without the networks of relations—families, colleagues, neighbours, and other human and non-human actors—that sustain life and labour. This perspective foregrounds the dense, relationality that makes both the worker and the work possible.

As the field of digital labour studies continues to expand, with scholars developing new methodologies, engaging with emerging technologies, and responding to the rapid transformation of technological development and deployment, I propose the concept of platform extractivism as a framework capable of addressing these complexities. This notion accounts for core-periphery relations in value exchange while remaining attentive to working conditions within platform-based and technologically mediated labour regimes. At the same time, it invites analysis beyond the strictly material toward the immaterial realms of knowledge, ideology, and meaning, and beyond the individual worker toward a broader constellation of social and non-social actors that indirectly shape how labour and production occur. In doing so, platform extractivism offers a unique understanding of how workers, work, and value are collectively produced within contemporary AI-driven economies.

## ACKNOWLEDGEMENTS

Research for this book was supported by the International Development Research Centre and the Social Science Research Council's Just Tech Fellowship, with funds provided by the Ford Foundation, MacArthur Foundation, and Surdna Foundation. The views, findings, conclusions, or recommendations expressed in this chapter are those of the author and do not necessarily represent those of the International Development Research Centre, the Social Science Research Council, Ford Foundation, MacArthur Foundation, and the Surdna Foundation or their governing boards.